\documentclass[12pt]{article}
\pdfoutput=1
\usepackage{putex}
\usepackage{graphicx}
\usepackage{caption}
\usepackage{subcaption}
\usepackage{epstopdf}
\usepackage{enumerate}
\usepackage{cite}
\usepackage{youngtab}
\usepackage{braket}
\usepackage{bbm}
\usepackage{tensor}
\usepackage{slashed}
\usepackage[aligntableaux=center]{ytableau}
\usepackage{multirow}

\newcommand{\abs}[1]{\left\lvert #1 \right\rvert}

\newcommand {\be} {\begin {equation}}
\newcommand {\ee} {\end {equation}}

\newcommand {\bes} {\begin {equation*}}
\newcommand {\ees} {\end {equation*}}

\newcommand{\es}[2] {\begin{equation} \label{#1} \begin{split} #2 \end{split} \end{equation}}

\newcommand{\cO}{{\cal O}}

\newcommand{\beq}{\begin{equation}}
\newcommand{\eeq}{\end{equation}}

\begin{document}

\preprint{PUPT-2553}

\institution{PU}{Joseph Henry Laboratories, Princeton University, Princeton, NJ 08544, USA}

\title{AdS$_4$/CFT$_3$ for Unprotected Operators}

\authors{Shai M.~Chester\worksat{\PU}}

\abstract{
We consider the four-point function of the lowest scalar in the stress-energy tensor multiplet in $\mathcal{N}=8$ ABJ(M) theory \cite{Aharony:2008ug, Aharony:2008gk}. At large central charge $c_T\sim N^{3/2}$, this correlator is given by the corresponding holographic correlation function in 11d supergravity on $AdS_4\times S^7$. We use Mellin space techniques to compute the leading $1/c_T$ correction to anomalous dimensions and OPE coefficients of operators that appear in this holographic correlator. For half and quarter-BPS operators, we find exact agreement with previously computed localization results. For the other BPS and non-BPS operators, our results match the $\mathcal{N}=8$ numerical bootstrap for ABJ(M) at large $c_T$, which provides a precise check of unprotected observables in AdS/CFT.
}
\date{\today}

\maketitle

\tableofcontents

\section{Introduction}
\label{intro}

The AdS$_4$/CFT$_3$ correspondence relates M-theory on $AdS_4\times S^7$ to certain 3d maximally supersymmetric ($\mathcal{N}=8$) superconformal field theories (SCFTs). These 3d SCFTs can all be described by a few infinite families of Chern-Simons (CS) theories with a product gauge group $G_1\times G_2$ coupled (in $\mathcal{N}=3$ notation) to two matter hypermultiplets transforming in the bifundamental representation. The ABJM$_{N,k}$ family \cite{Aharony:2008ug} has gauge group $U(N)_k\times U(N)_{-k}$, where the CS coupling $k=1,2$.\footnote{ABJM$_{1,1}$ is a free theory of 8 real scalars and 8 Majorana fermions, and for $N>1$ ABJM$_{N,1}$ is a product of this free theory and an interacting theory. We only consider the interacting sector in this work.} The ABJ$_{N}$ family \cite{Aharony:2008gk} has gauge group $U(N+1)_2\times U(N)_{-2}$, where $k$ is now fixed to 2 \footnote{A third family of $\mathcal{N}=8$ SCFTs are the BLG$_k$ \cite{Bagger:2007vi,Bagger:2007jr,Bagger:2006sk,Gustavsson:2007vu} theories with gauge group $SU(2)_{k}\times SU(2)_{-k}$, but for $k=1,2,3,4$ they are dual to certain ABJ(M) theories \cite{Lambert:2010ji,Bashkirov:2011pt,Agmon:2017lga}, while for $k>4$ they have no known M-theory interpretation.}. We refer collectively to both families as ABJ(M). These theories are conjectured to be effective theories on $N$ coincident M2-branes placed at a $\mathbb{C}^4/\mathbb{Z}_k$ singularity, so that when $N\to\infty$ they contain a sector described by weakly coupled supergravity on $AdS_4\times S^7$. It is convenient to parameterize these theories by the central charge $c_T$, which is defined as the coefficient of the canonically normalized stress tensor two point function \cite{Osborn:1993cr}
 \es{TmnCorr}{
  \langle T_{\mu\nu}(x) T_{\rho \sigma}(0) \rangle =  \frac{c_T}{64}\left( P_{\mu\rho} P_{\nu\sigma} + P_{\nu \rho} P_{\mu \sigma} - P_{\mu\nu} P_{\rho\sigma} \right) \frac{1}{16 \pi^2 x^2} \,,
 }
where $P_{\mu\nu} \equiv \eta_{\mu\nu} \nabla^2 - \partial_\mu \partial_\nu$ and $c_T=1$ for a real massless scalar or Majorana fermion. As $N\to\infty$, we have \cite{Chester:2014fya}
\es{largeNcT}{
c_T\approx\frac{64}{3\pi}\sqrt{2k}N^{3/2}\,,
}
so that the $c_T\to\infty$ limit of ABJ(M) is conjectured to describe weakly coupled supergravity.

The conjectured relation between ABJ(M) and M2-branes has been checked in several ways. The original authors \cite{Aharony:2008ug} matched the moduli spaces and chiral operators on each side. The index of chiral operators was computed and matched in \cite{Kim:2009wb}. The $S^3$ free energy was matched at leading order in $1/c_T$ \cite{Drukker:2010nc,Herzog:2010hf} and subsequently the logarithmic term \cite{Bhattacharyya:2012ye}. Aside from the logarithmic term, none of these matches have gone beyond leading order, though, nor have they matched any unprotected local CFT data (i.e. scaling dimensions and OPE coefficients). A major difficulty is that the IR fixed point of ABJ(M) is strongly coupled for all $c_T$, while at large $c_T$ it is difficult to compute even tree level four-point functions in weakly coupled supergravity on $AdS_4\times S^7$.

Progress was made recently in \cite{Zhou:2017zaw}, which computed the tree level supergravity contribution to the Mellin space \cite{Mack:2009mi} holographic four point function of the lowest scalar in the stress tensor multiplet.\footnote{For other recent progress on applications of Mellin space to CFTs, see for instance \cite{Penedones:2010ue,Costa:2012cb,Fitzpatrick:2011ia,Rastelli:2017ecj,Fitzpatrick:2011hu,Paulos:2011ie,Nandan:2011wc,Aharony:2016dwx,Yuan:2017vgp,Cardona:2017tsw,Rastelli:2016nze,Gopakumar:2016wkt,Gopakumar:2016cpb,Goncalves:2014rfa,Nizami:2016jgt,Paulos:2012nu,Rastelli:2017ymc,Rastelli:2017udc,Goncalves:2014ffa}.} This correlator was then expanded in terms of conformal blocks to read off the $1/c_T$ correction to the scaling dimension of the lowest unprotected operator that appears, which was matched to the large $c_T$ limit of ABJ(M) theory\footnote{At order $1/c_T$, the different ABJ(M) theories are indistinguishable.} computed from the $\mathcal{N}=8$ numerical bootstrap \cite{Agmon:2017xes}. 

We extend this result by extracting the $1/c_T$ corrections to the rest of the low-lying CFT data in this Mellin amplitude, including both BPS and non-BPS operators. Three new difficulties appear when considering operators other than the lowest unprotected operator. Firstly, the superconformal primary for a given multiplet may appear as a conformal primary in another multiplet, so expanding in conformal blocks is ambiguous. We resolve this problem by expanding in superconformal blocks, using the explicit expressions computed in \cite{Chester:2014fya}. Secondly, unlike in even dimensions, there is no closed form for the 3d conformal blocks, which makes it hard to extract CFT data from Mellin amplitudes for arbitrary twist and spin, as was done in even dimension \cite{Alday:2014tsa,Aharony:2016dwx,Heslop:2017sco}. In this work, we develop an efficient algorithm for extracting CFT data order by order in the twist, based on the expansion of 3d conformal blocks into lightcone blocks initiated in \cite{Hogervorst:2016hal,Simmons-Duffin:2016wlq}. Lastly, we expect there to be $n+1$ unprotected operators with the $n$th lowest twist, so for all but the lowest twist there will be mixing that cannot be resolved from studying the stress tensor four-point function alone. For these higher twist operators, our results should be interpreted as weighted averages, as we will explain further below.

After extracting the $1/c_T$ corrections to the CFT data, we compare to previously computed analytical and numerical results. The OPE coefficients of $\frac12$ and $\frac14$ BPS operators in the stress tensor four-point function were computed to all orders in $1/c_T$ in \cite{Agmon:2017xes} by applying matrix model techniques \cite{Marino:2011eh,Nosaka:2015iiw} to the 1d topological sector \cite{Chester:2014mea} of 3d $\mathcal{N}=8$ theories.\footnote{For other work on this sector see \cite{Beem:2016cbd,Beem:2013sza,Dedushenko:2017avn,Dedushenko:2016jxl}.} We find that the $1/c_T$ terms in these expressions exactly match our Mellin space calculation. For the other CFT data, we compare to the $\mathcal{N}=8$ numerical bootstrap results for ABJ(M) at large $c_T$ \cite{Agmon:2017xes}, and find a precise numerical match.\footnote{For higher twist unprotected operators, we can only make this comparison at large spin where the effects of the mixing are expected to be subleading.}

These matches constitute a precision test of AdS$_4$/CFT$_3$ for the following reasons. The Mellin space calculation relied on the assumption that supergravity has a standard two-derivative Einstein-Hilbert term and is equivalent to summing up the appropriate Witten diagrams, and so can be considered an AdS$_4$ calculation. The exact calculation of the $\frac12$ and $\frac14$ BPS OPE coefficients used the explicit form of the ABJ(M) lagrangian, and so is necessarily a CFT$_3$ calculation. The numerical bootstrap results were computed by assuming that ABJ(M) at large $c_T$ saturates the boundary of the allowed region of $\mathcal{N}=8$ theories, in which case it is expected to be the unique solution of the bootstrap equations. This assumption was motivated by observing that the allowed region was saturated at large $c_T$ by the known curves for the $\frac12$ and $\frac14$-BPS OPE coefficients, and so must also be considered a CFT$_3$ calculation.

The rest of this paper is organized as follows. In Section \ref{4point} we review the decomposition of the four-point function of the lowest scalar in the $\mathcal{N}=8$ stress-tensor multiplet. In Section \ref{leading}, we review the computation of the leading order CFT data in this correlator. In Section \ref{subleading}, we present an algorithm for extracting CFT data from maximally supersymmetric AdS$_4$ Mellin amplitudes, and apply it to the tree level amplitude computed in \cite{Zhou:2017zaw}. We then compare previously computed analytical and numerical results. In Section \ref{conclusion} we end with a discussion of our results and future directions. Appendix \ref{hog} reviews how to compute lightcone blocks from 3d blocks.

\section{Four-point function of stress-tensor}
\label{4point}

Let us begin by reviewing some general properties of the four-point function of the stress-tensor multiplet in an $\mathcal{N}=8$ SCFT, and of the constraints imposed by the $\mathfrak{osp}(8|4)$ superconformal algebra (for more details, the reader is referred to e.g. \cite{Minwalla:1997ka,Bhattacharya:2008zy,Dolan:2008vc}).

Unitary irreps of $\mathfrak{osp}(8|4)$ are specified by the quantum numbers of their bottom component, namely by its scaling dimension $\Delta$, Lorentz spin $j$, and $\mathfrak{so}(8)$ R-symmetry irrep with Dynkin labels $[a_1\, a_2 \, a_3 \, a_4]$, as well as by various shortening conditions.  There are twelve different types of multiplets that we list in Table~\ref{Multiplets}.\footnote{The convention we use in defining these multiplets is that the supercharges transform in the ${\bf 8}_v = [1000]$ irrep of $\mathfrak{so}(8)$.}  
\begin{table}[http]
\begin{center}
\begin{tabular}{|l|c|c|c|c|}
\hline
 Type     & BPS    & $\Delta$             & Spin & $\mathfrak{so}(8)$  \\
 \hline 
 $(A,0)$ (long)      & $0$    & $\ge \Delta_0 + j+1$ & $j$  & $[a_1 a_2 a_3 a_4]$  \\
 $(A, 1)$  & $1/16$ & $\Delta_0 + j +1$    & $j$  & $[a_1 a_2 a_3 a_4]$  \\
 $(A, 2)$  & $1/8$  & $\Delta_0 + j +1$    & $j$  & $[0 a_2 a_3 a_4]$   \\
 $(A, 3)$  & $3/16$  & $\Delta_0 + j +1$    & $j$  & $[0 0 a_3 a_4]$     \\
 $(A, +)$  & $1/4$  & $\Delta_0 + j +1$    & $j$  & $[0 0 a_3 0]$       \\
 $(A, -)$  & $1/4$  & $\Delta_0 + j +1$    & $j$  & $[0 0 0 a_4]$       \\
 $(B, 1)$  & $1/8$  & $\Delta_0$           & $0$  & $[a_1 a_2 a_3 a_4]$ \\
 $(B, 2)$  & $1/4$  & $\Delta_0$           & $0$  & $[0 a_2 a_3 a_4]$   \\
 $(B, 3)$  & $3/8$  & $\Delta_0$                  & $0$  & $[0 0 a_3 a_4]$     \\
 $(B, +)$  & $1/2$  & $\Delta_0$           & $0$  & $[0 0 a_3 0]$       \\
 $(B, -)$  & $1/2$  & $\Delta_0$           & $0$  & $[0 0 0 a_4]$       \\
 conserved & $5/16$  & $j+1$                & $j$  & $[0 0 0 0]$         \\
 \hline
\end{tabular}
\end{center}
\caption{Multiplets of $\mathfrak{osp}(8|4)$ and the quantum numbers of their corresponding superconformal primary operator. The conformal dimension $\Delta$ is written in terms of $\Delta_0 \equiv a_1 + a_2 + (a_3 + a_4)/2$.  The Lorentz spin can take the values $j=0, 1/2, 1, 3/2, \ldots$.  Representations of the $\mathfrak{so}(8)$ R-symmetry are given in terms of the four $\mathfrak{so}(8)$ Dynkin labels, which are non-negative integers.}
\label{Multiplets}
\end{table}

The stress-tensor multiplet is of $(B, +)$ type, and its superconformal primary has $\Delta=1$, $j=0$, and $\mathfrak{so}(8)$ irrep ${\bf 35}_c = [0020]$.  Let us denote this superconformal primary by $\cO_{\text{Stress}, IJ}(\vec{x})$.  (The indices here are ${\bf 8}_c$ indices, and  $\cO_{\text{Stress}, IJ}(\vec{x})$ is a rank-two traceless symmetric tensor.)  In order to not carry around the $\mathfrak{so}(8)$ indices, it is convenient to contract them with an auxiliary polarization vector $Y^I$ that is constrained to be null $Y \cdot Y \equiv \sum_{I=1}^8 (Y^I)^2 = 0$, thus defining 
 \es{ODef}{
  \cO_\text{Stress}(\vec{x},Y) \equiv \cO_{\text{Stress}, IJ}(\vec{x}) Y^I Y^J \,.
 }

In the rest of this paper we will only consider the four-point function of $\cO_\text{Stress}(\vec{x},Y)$. Superconformal invariance implies that it takes the form
 \es{FourPointO}{
  \langle \cO_\text{Stress}(\vec{x}_1,Y_1) \cO_\text{Stress}(\vec{x}_2,Y_2)
   &\cO_\text{Stress}(\vec{x}_3,Y_3) \cO_\text{Stress}(\vec{x}_4,Y_4) \rangle=\frac{( Y_1\cdot Y_2 )^2  ( Y_3\cdot Y_4 )^2 }
     {\abs{\vec{x}_{12}}^2\abs{\vec{x}_{34}}^2 }\mathcal{A}(U,V;\sigma,\tau) \,,\\
    &\qquad\qquad\qquad\;\;\,\mathcal{A}(U,V;\sigma,\tau)=
     \sum_{{\mathcal{M}}\, \in\, \mathfrak{osp}(8|4)} \lambda_{\mathcal{M}}^2 \mathcal{G}_{\mathcal{M}}(U, V; \sigma, \tau) \,,
 }
where 
 \es{uvsigmatauDefs}{
  U \equiv \frac{\vec{x}_{12}^2 \vec{x}_{34}^2}{\vec{x}_{13}^2 \vec{x}_{24}^2} \,, \qquad
   V \equiv \frac{\vec{x}_{14}^2 \vec{x}_{23}^2}{\vec{x}_{13}^2 \vec{x}_{24}^2}  \,, \qquad
   \sigma\equiv\frac{(Y_1\cdot Y_3)(Y_2\cdot Y_4)}{(Y_1\cdot Y_2)(Y_3\cdot Y_4)}\,,\qquad \tau\equiv\frac{(Y_1\cdot Y_4)(Y_2\cdot Y_3)}{(Y_1\cdot Y_2)(Y_3\cdot Y_4)} \,,
 }
$\mathcal{G}_{\mathcal{M}}$ are superconformal blocks, and $\lambda^2_{\mathcal{M}}$ are the OPE coefficients squared for each supermultiplet $\mathcal{M}$. In Table~\ref{opemult} we list the $\mathcal{M}$ that may appear in this four-point function, following the constraints discussed in \cite{Ferrara:2001uj}. Since these are the only multiplets we will consider in this paper, we denote the short multiplets other than the stress-tensor as $(B,+)$ and $(B,2)$, the semi-short multiplets as $(A,2)_j$ and $(A,+)_j$ where $j$ is the spin, and the long multiplet as $(A,0)_{j,n,q}$, where $n=0,1,\dots$ denotes the leading order twist $2n+2$ and $q=0\,,\dots n$ denotes the distinct operators with the same leading order quantum numbers.

\begin{table}
\centering
\begin{tabular}{|c|c|r|c|c|}
\hline
Type    & $(\Delta,j)$     & $\mathfrak{so}(8)$ irrep  &spin $j$ & Name \\
\hline
$(B,+)$ &  $(2,0)$         & ${\bf 294}_c = [0040]$& $0$ & $(B, +)$ \\ 
$(B,2)$ &  $(2,0)$         & ${\bf 300} = [0200]$& $0$ & $(B, 2)$ \\
$(B,+)$ &  $(1,0)$         & ${\bf 35}_c = [0020]$ & $0$ & Stress \\
$(A,+)$ &  $(j+2,j)$       & ${\bf 35}_c = [0020]$ &even & $(A,+)_j$ \\
$(A,2)$ &  $(j+2,j)$       & ${\bf 28} = [0100]$ & odd & $(A,2)_j$ \\
$(A,0)$ &  $\Delta\ge j+1$ & ${\bf 1} = [0000]$ & even & $(A,0)_{j,n,q}$\\
\hline
\end{tabular}
\caption{The possible superconformal multiplets in the $\cO_\text{Stress} \times  \cO_\text{Stress}$ OPE\@.  The $\mathfrak{so}(3, 2) \oplus \mathfrak{so}(8)$ quantum numbers are those of the superconformal primary in each multiplet.}
\label{opemult}
\end{table}

Of particular importance will be the OPE coefficient for the stress-tensor multiplet. In the conventions of \cite{Chester:2014fya}, if we normalize ${\cal O}_\text{Stress}$ such that the OPE coefficient of the identity operator is $\lambda_{\text{Id}} = 1$, then
 \es{cTRel}{
  \lambda_{\text{Stress}}^2 = \frac{256}{c_T} \,,
 }
where $c_T$ is the coefficient appearing in the two-point function \eqref{TmnCorr} of the canonically normalized stress tensor.

Each superconformal block $\mathcal{G}_{\mathcal{M}}$ receives contributions from conformal primaries $(\Delta, j)_{[0a_1 a_20]}$ with different spins $j$, scaling dimensions $\Delta$, and irreps $[0\, (a-b)\, (2b)\, 0]$ for $a=0,1,2$ and $b=0,\dots a$ that appear in $[0020]\otimes [0020]$. These conformal primaries can be found by decomposing $\mathfrak{osp}(8|4)$ characters \cite{Dolan:2008vc} into characters of the maximal bosonic sub-algebra $\mathfrak{so}(3,2)\oplus\mathfrak{so}(8)$. This decomposition was performed in \cite{Chester:2014fya}. For instance, the conformal primaries that can contribute to $\mathcal{G}_\text{Stress}$ are given in Table \ref{Bp}. For the other $\mathcal{G}_\mathcal{M}$, see Table $4-8$ in \cite{Chester:2014fya}.

\begin{table}[htpb]
\centering
\begin{tabular}{|l||c|c|c|c|}
\hline 
 spin: & \multicolumn{3}{c|}{\text{dimension}}  \\
\hline
$\mathfrak{so}(8)$ irrep& 1&2&3\\         
\hline\hline
{$[0000]=\bold{1}$}       & --          & --    &  $ {2}$   \\
{$[0100]=\bold{28}$}      & --          & ${1} $    & --\\
$[0020]=\bold{35}_c$   & {0}         & --       &--\\
\hline
\end{tabular}
\caption{All possible conformal primaries in $\cO_\text{Stress}\times \cO_\text{Stress}$ corresponding to the stress-tensor superconformal multiplet.}\label{Bp}
\end{table}

The superconformal block $\mathcal{G}_{\mathcal{M}_{\Delta,j}}$ whose superprimary has dimension $\Delta$ and spin $j$ can then be decomposed into conformal blocks $G_{\Delta',j'}$ of the conformal primaries in $\mathcal{M}$ as
 \es{GExpansion}{
  {\cal G}_{\mathcal{M}_{\Delta,j}}(U, V; \sigma, \tau) = \sum_{a=0}^2 \sum_{b = 0}^a Y_{ab}(\sigma, \tau)  \sum_{(\Delta',j')\in\mathcal{M}} A^\mathcal{M}_{ab \Delta' j'}(\Delta, j) G_{\Delta',j'}(U,V) \,,
 }  
where the quadratic polynomials $Y_{ab}(\sigma, \tau)$ are eigenfunctions of the $\mathfrak{so}(8)$ Casimir, and are given in \cite{Dolan:2003hv,Nirschl:2004pa} as
 \es{polyns}{
   Y_{00}(\sigma, \tau) &= 1 \,, \\
   Y_{10}(\sigma, \tau) &= \sigma - \tau \,, \\
   Y_{11}(\sigma, \tau) &= \sigma + \tau -\frac{1}{4} \,, \\
   Y_{20}(\sigma, \tau) &= \sigma^2 + \tau^2 - 2\sigma\tau - \frac{1}{3}(\sigma + \tau) + \frac{1}{21} \,,\\
   Y_{21}(\sigma, \tau) &= \sigma^2 - \tau^2 - \frac{2}{5}(\sigma - \tau) \,,\\
   Y_{22}(\sigma, \tau) &= \sigma^2 + \tau^2 + 4\sigma\tau - \frac{2}{3}(\sigma+\tau) + \frac{1}{15} \,.
 }
  The $A^\mathcal{M}_{ab\Delta'j'}(\Delta, j)$ are rational function of $\Delta$ and $j$ that were computed in \cite{Chester:2014fya} using the superconformal Ward identity derived in \cite{Dolan:2004mu}. For instance, for $\mathcal{G}_\text{Stress}$ we have
\es{stressBlock}{
A^\text{Stress}_{1110} = 1 \,,\qquad A^\text{Stress}_{1021} = -1\,, \qquad A^\text{Stress}_{0032} = \frac{1}{4} \,,
}
which corresponds to the conformal primaries in Table \ref{Bp}. For the other $\mathcal{G}_\mathcal{M}$, see Appendix C in  \cite{Chester:2014fya}.

\section{$c_T\to\infty$ theory}
\label{leading}

We will now compute the CFT data in the $\cO_\text{Stress}$ four-point function in a $1/c_T$ expansion. We expect scaling dimensions of the unprotected $(A,0)_{j,n,q}$ and OPE coefficients of all operators to get corrections as
\es{corrections}{
\Delta_{\mathcal{M}}&=\Delta^{(0)}_{\mathcal{M}}+c_T^{-1}\Delta_{\mathcal{M}}^{(1)}+\dots\,,\\
 \lambda^2_\mathcal{M}&=a_\mathcal{M}^{(0)}+c_T^{-1}a_\mathcal{M}^{(1)}+\dots\,.
}
 We begin by reviewing the strict $c_T\to\infty$ limit. From the AdS$_4$ perspective, this limit corresponds to classical supergravity on $AdS_4\times S^7$, so the stress tensor four-point amplitude is given by disconnected Witten diagrams, which contribute only to double trace operators like $[\cO_\text{Stress}\cO_\text{Stress}]$.

From the CFT$_3$ perspective, this limit corresponds to a generalized free field theory (GFFT) generated by the dimension one operator $\cO_\text{Stress}$. The $\cO_\text{Stress}$ four-point function can be computed from Wick contractions using the two-point function $\langle {\cal O}_\text{Stress}(\vec{x}, Y_1)  {\cal O}_\text{Stress}(0, Y_2) \rangle = \frac{(Y_1 \cdot Y_2)^2}{\abs{x}^2}$, which gives the leading order conformal block expansion
\es{mean4}{
 \mathcal{A}^{(0)}(U,V;\sigma,\tau)=
1+U\sigma^2+\frac{U}{V}\tau^2\,.
}
We can now determine the leading order scaling dimensions $\Delta^{(0)}_\mathcal{M}$ and OPE coefficients squared $a_\mathcal{M}^{(0)}$ by expanding \eqref{mean4} into superconformal blocks $\mathcal{G}_\mathcal{M}(U,V,\sigma,\tau)$ and then comparing to \eqref{FourPointO}. To perform this expansion it is convenient to use the $r$ and $\eta$ variables defined in \cite{Hogervorst:2013sma} as
\es{reta}{
 U\equiv\frac{16r^2}{\left(1+r^2+2r\eta\right)^2}\,,\quad V\equiv \frac{\left(1+r^2-2r\eta\right)^2}{\left(1+r^2+2r\eta\right)^2}\,.
}
The advantage of the $r$ and $\eta$ variables is that in the $r\to0$ limit for fixed $\eta$, the conformal blocks $G_{\Delta,j}(r,\eta)$ can be organized according to their scaling dimension as
\es{confBlock}{
G_{\Delta,j}=r^\Delta P_{j}(\eta)+O(r^{\Delta+1})\,,
}
where the higher orders in $r$ can be found, for instance, in \cite{Kos:2013tga}. Using this expansion, and the explicit definitions of $\mathcal{G}_\mathcal{M}(r,\eta,\sigma,\tau)$ in terms of $G_{\Delta,j}(r,\eta)$ given in \eqref{GExpansion}, we can efficiently read off the leading order in $1/c_T$ OPE coefficients listed in Table \ref{Avalues}, as well as the leading order scaling dimensions for the unprotected operator $(A,0)_{j,n,q}$, which takes the form
\es{superMean}{
\Delta_{(A,0)_{j,n,q}}^{(0)}=2+j+2n\,.
}
Note that the CFT data for $(A,0)_{j,n,q}$ does not depend on $q$ to this order. 

 \begin{table}[htp]
\begin{center}
\begin{tabular}{|l|c|}
\hline
 \multicolumn{1}{|c|}{Type $\cal M$}  & $c_T\to\infty$ OPE coefficient squared $a_\mathcal{M}^{(0)}$ \\
  \hline
  $(B,2)$  & $\quad\qquad\qquad\quad\qquad\qquad\qquad\quad\;\,32/3\approx10.667$\\
  $(B,+)$ & $\;\;\quad\qquad\qquad\qquad\qquad\qquad\quad\;\,16/3\approx5.333$\\
  $(A,2)_1$  & $\;\;\quad\qquad\qquad\qquad\qquad\quad\;\,1024 / 105\approx9.752$\\
  $(A,2)_3$  & $\;\;\qquad\qquad\qquad\qquad\quad\;\,131072 / 8085\approx16.212$\\
  $(A,2)_5$  & $\qquad\qquad\qquad\quad\;\,33554432 / 1486485\approx22.573$\\
  $(A,+)_0$  & $\;\;\quad\qquad\qquad\qquad\qquad\qquad\quad\;\,64 / 9\approx7.111$\\
  $(A,+)_2$  & $\quad\qquad\qquad\qquad\qquad\quad\;\,16384 / 1225\approx13.375$\\
  $(A,+)_4$ & $\;\;\quad\qquad\qquad\qquad\quad\;\,1048576 / 53361\approx19.651$\\
  $(A,+)_6$ & $\;\;\quad\qquad\qquad\;\,1073741824/41409225\approx25.930$\\
  $(A,0)_{0,0,0}$ &  $\quad\qquad\qquad\qquad\qquad\qquad \quad\;\;32 / 35\approx0.911$\\
  $(A,0)_{2,0,0}$ &  $\;\;\quad\qquad\qquad\qquad\qquad\quad\; \,2048 / 693\approx2.955$\\
  $(A,0)_{4,0,0}$ & $ \;\;\qquad\qquad\qquad\quad\;\,1048576/225225\approx4.656$\\
  $(A,0)_{6,0,0}$ & $ \;\;\qquad\quad\qquad\quad67108864/10669659\approx6.290$\\
  $(A,0)_{8,0,0}$ & $ \;\;\qquad\quad\;\;34359738368/4350310965\approx7.899$\\
  $(A,0)_{10,0,0}$ & $ \;\qquad\,2199023255552/231618204675\approx9.494$\\
  $(A,0)_{12,0,0}$ & $ 2251799813685248/203176892887605\approx11.083$\\
  $(A,0)_{0,1,q}$ &  $\quad\qquad\qquad \qquad\qquad\;\,\qquad\;\,256/693\approx0.369$\\
  $(A,0)_{2,1,q}$ &  $\quad\qquad\qquad \qquad\;\;\qquad\;\,65536/45045\approx1.455$\\
  $(A,0)_{4,1,q}$ &  $\quad\qquad\qquad\;\; \qquad\;\,8388608/3556553\approx2.359$\\
  $(A,0)_{6,1,q}$ &  $\quad\qquad\qquad\;\, 1073741824/334639305\approx3.209$\\
  $(A,0)_{8,1,q}$ &  $\quad \qquad\;\,274877906944/68123001375\approx4.035$\\
  $(A,0)_{10,1,q}$ & $\;\,140737488355328/29025270412515\approx4.849$\\
  \hline
\end{tabular}
\end{center}
\caption{Values of OPE coefficients squared $a_\mathcal{M}^{(0)}$ at $c_T\to\infty$ for low-lying multiplets $\mathcal{M}$ in $\cO_\text{Stress}\times\cO_\text{Stress}$.}\label{Avalues}
\end{table}

\section{$1/c_T$ corrections}
\label{subleading}

We now compute the $1/c_T$ correction to OPE coefficients and scaling dimensions of operators in the $\cO_\text{Stress}$ four point function. From the CFT$_3$ perspective, the only quantities that have been computed analytically to this order are the short operator OPE coefficients $\lambda_{(B,2)}$ and $\lambda_{(B,+)}$, which are known to all orders in $1/c_T$. The other CFT data has been computed numerically for all $c_T$ using the conformal bootstrap in \cite{Agmon:2017xes}. We will describe these CFT results in more detail in Section \ref{comparison}, when we compare them to the AdS$_4$ results.

From the AdS$_4$ perspective, the $1/c_T$ correction corresponds to tree level supergravity on $AdS_4\times S^7$, which is dual to all $\mathcal{N}=8$ ABJ(M) theories at this order. The tree level $\cO_\text{Stress}$ four-point function receives contributions from contact and exchange Witten diagrams of single trace operators. This correlator was computed explicitly in Mellin space in \cite{Zhou:2017zaw}. We will now review this Mellin space amplitude, and then extract all the relevant CFT data from it.

\subsection{Mellin space amplitude}
\label{mellin}

The connected Mellin space amplitude $M(s,t)$ for four identical scalars with scaling dimension $\Delta$ is defined in terms of the connected conformal block expansion $\mathcal{A}$ defined in \eqref{FourPointO} as
\es{mellinDef}{
\mathcal{A}(U,V;\sigma,\tau)=\int_{-i\infty}^{i\infty}\frac{ds dt}{(4\pi i)^2} U^{\frac s2}V^{\frac t2-\Delta}M(s,t,\sigma,\tau)\Gamma^2\left[\Delta-\frac t2\right]\Gamma^2\left[\Delta-\frac s2\right]\Gamma^2\left[\Delta-\frac u2\right]\,,
}
where the Mellin space variables satisfy the constraint $s+t+u=4\Delta$. The two integration contours run parallel to the imaginary axis, such that all poles of the Gamma functions are on one side or the other of the contour. The poles of the Gamma functions precisely capture the contribution of double trace operators in the OPE. The contributions from the single trace exchange and the contact diagrams were computed by \cite{Zhou:2017zaw} using the superconformal Ward identities \cite{Dolan:2004mu} and the assumption that the contact diagram is linear in $s,t,u$, which is implied by the two derivative Einstein-Hilbert term in the supergravity action. The resulting Mellin amplitude takes the form
\es{Xinan}{
M(s,t;\sigma,\tau)=M_\text{$s$-exchange}+M_\text{$t$-exchange}+M_\text{$u$-exchange}+M_\text{contact}\,,
}
where the contact term is
\es{Xcontact}{
M_\text{contact}=-\frac{\pi\lambda_s}{2}\left(s+u\sigma^2+t\tau^2-4(t+u)\sigma\tau-4(s+u)\sigma-4(s+t)\tau\right)\,,
}
and the $s$-channel exchange term receives contributions from the graviton, vector, and scalar components of the graviton multiplet as
\es{sChannel}{
M_\text{$s$-exchange}&=\lambda_s\left[\frac13M_\text{graviton}-4(\sigma-\tau)M_\text{vector}+4\left(\sigma+\tau-\frac14\right)M_\text{scalar}\right]\,,\\
M_\text{graviton}&=\sum_{m=0}^\infty\frac{3\sqrt{\pi}(-1)^m \Gamma[-\frac32-m]}{4m!\Gamma[\frac12-m]^2}\frac{4m^2-8ms+8m+4s^2+8st-20s+8t^2-32t+35}{s-(2m+1)}\,,\\
M_\text{vector}&=\sum_{m=0}^\infty\frac{\sqrt{\pi}(-1)^m}{(2m+1)m!\Gamma[\frac12-m]}\frac{2t+s-4}{s-(2m+1)}\,,\\
M_\text{scalar}&=\sum_{m=0}^\infty\frac{\sqrt{\pi}(-1)^m }{m!\Gamma[\frac12-m]}\frac{1}{s-(2m+1)}\,.\\
}
The $t$- and $u$-channel exchange is related to the $s$-channel exchange by crossing symmetry
\es{tandu}{
M_\text{$t$-exchange}(s,t;\sigma,\tau)&=\tau^2 M_\text{$s$-exchange}(t,s;\sigma/\tau,1/\tau)\,,\\
M_\text{$u$-exchange}(s,t;\sigma,\tau)&=\sigma^2 M_\text{$s$-exchange}(u,t;1/\sigma,\tau/\sigma)\,.
}
Finally, the overall coefficient $\lambda_s$ is normalized in terms of $\lambda^2_\text{Stress}$. In \cite{Zhou:2017zaw}, this was written for ABJM$_{N,1}$ as
\es{lambdas}{
\lambda_s=-\frac{3\sqrt{2}}{4\pi^2 N^{3/2}}\,.
}
Comparing this to our large $N$ expression for $c_T$ in \eqref{largeNcT}, we get
\es{lambdascT}{
\lambda_s=-\frac{32}{\pi^3 c_T}\,,
}
which completes the description of the tree level Mellin amplitude $M(s,t;\sigma,\tau)$.

\subsection{Extracting CFT data}
\label{extract}

To extract OPE coefficients and scaling dimensions from the tree level amplitude, we need to expand the conformal blocks in $\mathcal{A}$ defined in \eqref{FourPointO} to order $1/c_T$. Using the expansion of the CFT data in \eqref{corrections}, we find that the tree level coefficient in $\mathcal{A}=\mathcal{A}^{(0)}+c_T^{-1}\mathcal{A}^{(1)}$ is 
\es{Aexpansion}{
\mathcal{A}^{(1)}(U,V;\sigma,\tau)=\sum_{\mathcal{M}_{\Delta,j}\in\mathfrak{osp}(8|4)} \left[ a_\mathcal{M}^{(1)} \mathcal{G}_\mathcal{M}(U,V;\sigma,\tau)+ a_{\mathcal{M}}^{(0)} \Delta_\mathcal{M}^{(1)}\partial_{\Delta}\mathcal{G}_{\mathcal{M}}(U,V;\sigma,\tau)\right]_{\Delta^{(0)}_\mathcal{M}} \,,
}
where the subscript ${\Delta^{(0)}_\mathcal{M}} $ denotes that the blocks for the unprotected operators should be evaluated with the leading order scaling dimension.

If we had an explicit position space expression for $\mathcal{A}^{(1)}(U,V;\sigma,\tau)$, as we had for the leading order $\mathcal{A}^{(0)}(U,V;\sigma,\tau)$ in \eqref{mean4}, then we could simply expand in $r$ and $\eta$ variables as described in Section \ref{leading}. The integrals in the Mellin transform \eqref{mellinDef} that relate $M(s,t;\sigma,\tau)$ to $\mathcal{A}(U,V;\sigma,\tau)$ cannot be performed for arbitrary $U$ and $V$, however, so we cannot obtain the CFT data using the expansion in $r$ of Section \ref{leading}. Instead, as is standard in the Mellin space literature, we use the lightcone expansion $U\ll1$ for fixed $V$. The conformal blocks are expanded as
\es{lightBlocksExp}{
G_{\Delta,j}(U,V)=\sum_{k=0}^\infty U^{\frac{\Delta-j}{2}+k}g_{\Delta,j}^{[k]}(V)\,,
}
where the lightcone blocks $g_{\Delta,j}^{[k]}(V)$ are labeled by the $k$-th lowest twist, and are only functions of $V$. They can be computed by decomposing 3d conformal blocks to 2d \cite{Hogervorst:2016hal}, which we review in Appendix \ref{hog}, and the answer can always be written as a finite sum of ${}_2F_1$ hypergeometric function. For instance, for $k=0,1$ the lightcone blocks are
\es{lightconeBlock}{
g_{\Delta,j}^{[0]}(V)&=\frac{\Gamma(j+1/2)}{4^\Delta\sqrt{\pi}j!}(1-V)^j \,{}_2F_1\left(\frac{\Delta+j}{2},\frac{\Delta+j}{2},\Delta+j,1-V\right)\,,\\
g_{\Delta,j}^{[1]}(V)&=\frac{\Gamma(j+1/2)(1-V)^{j-2}}{2(2j-1)(2\Delta-1) 4^\Delta\sqrt{\pi}j! }\left[
2(j+\Delta)(j+\Delta-2j \Delta)\, {}_2F_1\left(\frac{\Delta+j-2}{2},\frac{\Delta+j}{2},\Delta+j,1-V\right)\right.\\
&\left.-(1+V)( \Delta^2+j^2(2\Delta-1)-2j(\Delta^2+\Delta-1) ) \,{}_2F_1\left(\frac{\Delta+j}{2},\frac{\Delta+j}{2},\Delta+j,1-V\right)
\right]
\,.\\
}
Using \eqref{lightBlocksExp} and the expansion \eqref{GExpansion} of superconformal blocks into conformal blocks, we can expand $\mathcal{A}^{(1)}$ in \eqref{Aexpansion} for $U\ll1$ as
\es{Aexpansion2}{
&\mathcal{A}^{(1)}(U,V;\sigma,\tau)= \sum_{a=0}^2 \sum_{b = 0}^a Y_{ab}(\sigma, \tau)  \sum_{\mathcal{M}_{\Delta,j}\in\mathfrak{osp}(8|4)} \sum_{(\Delta',j')\in\mathcal{M}}\sum_{k=0}^\infty U^{\frac{\Delta'-j'}{2}+k} \\ &\left[  a_\mathcal{M}^{(1)} A^\mathcal{M}_{ab \Delta' j'}(\Delta, j) g^{[k]}_{\Delta',j'}(V)
 +a_{\mathcal{M}}^{(0)}   \Delta^{(1)}_\mathcal{M}\left[\partial_{\Delta}+\frac{\log U}{2}\right]\left[A^\mathcal{M}_{ab \Delta' j'}(\Delta, j) g^{[k]}_{\Delta',j'}(V)\right]\right]_{\Delta^{(0)}_\mathcal{M}}\,.
}
Note that for the unprotected $(A,0)_{j,n,q}$, the conformal primary scaling dimensions $\Delta'$ are shifts of the superconformal primary scaling dimension $\Delta$, so the $\Delta$-derivative will act on these conformal blocks as well as their coefficients $A^\mathcal{M}_{ab \Delta' j'}(\Delta, j) $. The utility of the lightcone expansion is that the $U$-dependence corresponds to the twist $\Delta-j$ of a conformal primary, and the $\log U$ term distinguishes between the scaling dimension and the OPE coefficient of that primary. In the Mellin transform \eqref{mellinDef}, one can isolate the $U^{\frac{\Delta'-j'}{2}+k}$ factor by taking the residue of the pole $s=\Delta'-j'+2k$. The $t$-integral can then be performed by summing all the poles, which yields a function of $V$. 

We can then extract the coefficients of a set of lightcone block using the orthogonality relationship for hypergeometric functions \cite{Heemskerk:2009pn}
\es{ortho}{
\delta_{p,p'}&=-\oint_{V=1} \frac{dV}{2\pi i}(1-V)^{p-p'-1} F_p(1-V) F_{1-p'}(1-V)\,,\\
F_p(x)&\equiv {}_2 F_1(p,p,2p,x)\,,
}
where we choose a contour that only contains the pole $V=1$. For instance, if we multiply $\mathcal{A}(U,V;\sigma,\tau)$ by $-(1-V)^{-1-j'} F_{1-\frac{\Delta'+j'}{2}}(1-V)$ and take the residue at $V=1$, then we will pick out all lightcone blocks $g^{[k]}_{\Delta,j}(V)$ with $j=j',j'+2,\dots, j'+2k$, as well as all $\partial_\Delta g^{[k]}_{\Delta,j}(V)$ with $j<j'+2k-1$. Combined with our ability to select the twist $
\Delta-j$ and $R$-symmetry structure $Y_{ab}(\sigma,\tau)$, as well as our knowledge of how each conformal primary contributes to the superconformal multiplet, this is enough to recursively solve for all $\Delta^{(1)}_\mathcal{M}$ and $a^{(1)}_\mathcal{M}$ for each superconformal multiplet $\mathcal{M}_{\Delta,j}$ using the following algorithm:

\begin{enumerate}
\item For a given supermultiplet $\mathcal{M}$, pick a conformal primary $(\Delta,j)_{[0a_1a_20]}$ from the tables $4-8$ in  \cite{Chester:2014fya}, and project the Mellin amplitude $M(s,t;\sigma,\tau)$ given in \eqref{Xinan} to this irrep by solving for the coefficient of $Y_{a_1+\frac{a_2}{2}\,\frac{a_2}{2}}(\sigma,\tau)$ as defined in \eqref{polyns}. 
\item Take the residue of the pole $s=\Delta-j$ in the Mellin transform \eqref{mellinDef}, leaving the sum over $m$ in \eqref{sChannel} unevaluated, then take the coefficient of $U^{\frac{\Delta-j}{2}}$ or $U^{\frac{\Delta-j}{2}}\log U$ to extract $a^{(1)}_\mathcal{M}$ or $\Delta^{(1)}_\mathcal{M}$, respectively, for all conformal primaries in $[0a_1a_20]$ with twist $\Delta-j$.
\item Compute the remaining $t$-integral in \eqref{mellinDef} by summing all poles with $t>0$.
\item Compare to \eqref{Aexpansion2}, multiply by $-(1-V)^{-1-j} F_{1-\frac{\Delta+j}{2}}(1-V)$, and perform the contour integral in \eqref{ortho} by taking the residue at $V=1$ to isolate all $g_{\Delta',j'}^{[k]}(V)$ with $j'=j,j+2,\dots, j+2k$ and $\partial_\Delta g^{[k]}_{\Delta',j'}(V)$ with $j'<j+2k-1$ in \eqref{Aexpansion2}, which includes the block $g_{\Delta,j}^{[0]}(V)$ for the desired conformal primary.
\item Perform the convergent infinite sum over $m$, from the exchange terms in \eqref{sChannel}.
\item Solve for the desired CFT data from \eqref{Aexpansion2} using the explicit expressions for $A^\mathcal{M}_{ab \Delta j}$ in Appendix C of \cite{Chester:2014fya} and subtract any other CFT data with the same $U$-dependence that remained after the $V$-integral in Step $4$.
\end{enumerate}
We will now demonstrate this algorithm in a series of increasingly more complicated examples for low-lying CFT data in the stress tensor four-point function.

\subsubsection{$a_{(B,2)}^{(1)}$ and $a_{(B,+)}^{(1)}$}
\label{short}
We begin with the short multiplets $(B,+)$ and $(B,2)$. For $(B,+)$, we choose the conformal primary $(2,0)_{[0040]}$, which happens to be the superconformal primary. This is a convenient choice, because it is the only conformal primary in any $\mathcal{M}$ with these quantum numbers, unlike e.g. $(3,1)_{[0120]}$ which appears in $(B,2)$ and $(A,+)_0$. We now take the residue of the pole $s=2$ in \eqref{mellinDef}, and find that the coefficient of $U Y_{22}$ in $\mathcal{A}(U,V;\sigma,\tau)$ is
\es{Bplus}{
&\mathcal{A}\big\vert_{UY_{22}}[V]=\int \frac{dt}{2\pi i}\frac{8}{c_T}\csc\left[\frac{\pi t}{2}\right]^2 V^{\frac t2-1}\left[\left(1-2\gamma-2\psi(t/2)\right)-\sum_{m=0}^\infty\left[\frac{32\pi^{-\frac12}(-1)^m}{3\Gamma[-m-3/2]m!}\right.\right.\\
&\left.\left.(3+4m(2+m))^{-2}\left(3+\frac{4}{1+2m-t}+\frac{4}{-1+2m+t}-4(1+m)(\gamma+\psi(t/2))\right)
\right]
\right]\,,
}
where $\gamma$ is the Euler-Mascheroni constant and $\psi$ is the Digamma function. This expression has $t>0$ poles for $t\in2\mathbb{Z}^+$, and $t=2m+1$ in the sum. We sum the residues from these poles, and then multiply by $\frac{F_{0}(1-V)}{V-1}=\frac{1}{V-1}$ and take the residue at $V=1$ to get
\es{Bplus2}{
\oint_{V=1}\frac{dV}{2\pi i}\frac{\mathcal{A}\big\vert_{UY_{22}}[V]}{V-1}&=\frac{1}{c_T}\left[-\frac{48}{\pi^2}+\sum_{m=0}^\infty\frac{(-1)^m512\left(7+4m+4\psi^{(1)}\left(\frac12+m\right)\right)}{3\pi^{\frac52}(3+4m(2+m))^2\Gamma\left[-\frac{3}{2}-m\right]m!}\right]\\
&=\frac{64}{c_T}\left(\frac{1}{9}+\frac{1}{3\pi^2}\right)\,.
}
From the block expansion for $\mathcal{A}(U,V;\sigma,\tau)$ in \eqref{Aexpansion2}, we see that integrating against $\frac{1}{V-1}$ and taking the coefficient of $U Y_{22}$ isolates the term $\frac{\lambda^2_{(B,+)}}{16}$, where $A^{(B,+)}_{2220}(2,0)=1$ because we chose the superconformal primary. We thus find
\es{BplusFinal}{
a_{(B,+)}^{(1)}=\frac{1024}{c_T}\left(\frac{1}{9}+\frac{1}{3\pi^2}\right)\,.
}
Performing the analogous calculation for $(B,2)$, by choosing the superconformal primary $(2,0)_{[0200]}$, which is also the only the only conformal primary in any $\mathcal{M}$ with these quantum numbers, yields
\es{B2Final}{
a_{(B,2)}^{(1)}=\frac{1024}{c_T}\left(-\frac{4}{9}+\frac{5}{3\pi^2}\right)\,.
}

\subsubsection{$a_{(A,+)_j}^{(1)}$ for $j=0,2,4,6$ and $a_{(A,2)_j}^{(1)}$ for $j=1,3,5$}
\label{semi-shorts}

For the semi-short operator $(A,+)_j$, we choose the conformal primary $(j+4,j+2)_{[0040]}$. Note that this is not the superconformal primary, but it has the advantage of being the only conformal primary in $\mathcal{M}$ with these quantum numbers for any $j$. If we had chosen the superconformal primary $(j+2,j)_{[0020]}$, then for $j=2$ this primary would have appeared in both $(A,+)_0$ and $(A,+)_2$. Another advantage of $(j+4,j+2)_{[0040]}$ is that it has the same twist and irrep as the conformal primary $(2,0)_{[0040]}$ that we chose for $(B,+)$, so we can use the same expression $\mathcal{A}\big\vert_{UY_{22}}[V]$ that was computed in \eqref{Bplus}. We now extract $g^{[0]}_{j+4,j+2}(V)$ by integrating with $\frac{F_{-j-2}(1-V)}{(V-1)^{j+3}}$, and perform the sum in $m$ to find
\es{Aplus2}{
\oint_{V=1}\frac{dV}{2\pi i}\frac{F_{-j-2}(1-V)}{(V-1)^{j+3}}\mathcal{A}\big\vert_{UY_{22}}[V]&=\begin{cases}
-\frac{160}{27} + \frac{560}{9 \pi^2}\qquad\quad\;\;\, \qquad j=0\\
-\frac{608}{315} +\frac{2596}{135 \pi^2}\qquad\quad\qquad j=2\\
-\frac{656}{2079} + \frac{44278}{14175 \pi^2}\quad\;\;\;\qquad j=4\\
-\frac{2272}{57915} + \frac{82517779}{212837625 \pi^2}\quad\;\;\; \;j=6\\
\end{cases}\,.
}
From the block expansion \eqref{Aexpansion2} we find
\es{Aplus3}{
\oint_{V=1}\frac{dV}{2\pi i}\frac{F_{-j-2}(1-V)}{(V-1)^{j+3}}\mathcal{A}\big\vert_{UY_{22}}[V]&=a^{(1)}_{(A,+)_j}\frac{ \Gamma(j+5/2)}{ 4^{j+2}3\sqrt{\pi}(j+2)!}\,,
}
where we used $A^{(A,+)_j}_{22\, j+4\,j+2}(j+2,j)=\frac{16}{3}$. Comparing to \eqref{Aplus2} we get
\es{AplusFinal}{
a_{(A,+)_0}^{(1)}&=-\frac{20480}{27} + \frac{71680}{9 \pi^2}\,,\\
a_{(A,+)_2}^{(1)}&=-\frac{19922944}{3675} + \frac{85065728}{1575 \pi^2}\,,\\
a_{(A,+)_4}^{(1)}&=-\frac{2751463424}{160083} + \frac{185715392512}{1091475 \pi^2}\,,\\
a_{(A,+)_6}^{(1)}&=-\frac{4879082848256}{124227675} + \frac{177205581071777792}{456536705625 \pi^2}\,.
}

The calculation for $(A,2)_j$ is more subtle, because there is no longer a twist 2 conformal primary that only appears in $(A,2)_j$. We choose the conformal primary $(j+4,j+2)_{[0120]}$, because it overlaps with fewer multiplets than other choices. Performing the usual first few steps, we find
\es{A22}{
\oint_{V=1}\frac{dV}{2\pi i}\frac{F_{-j-2}(1-V)}{(V-1)^{j+3}}\mathcal{A}\big\vert_{UY_{21}}[V]&=\begin{cases}
\frac{208}{15} - \frac{6104}{45 \pi^2}\qquad\quad\;\;\,  j=1\\
\frac{496}{189} - \frac{366278}{14175 \pi^2}\qquad\;\; j=3\\
\frac{152}{429} - \frac{5507939}{1576575 \pi^2}\quad\;\;\; j=5\\
\end{cases}\,.
}
From the block expansion \eqref{Aexpansion2} and the tables in \cite{Chester:2014fya} we find
\es{A23}{
&\oint_{V=1}\frac{dV}{2\pi i}\frac{F_{-j-2}(1-V)}{(V-1)^{j+3}}\mathcal{A}\big\vert_{UY_{21}}[V]=\\
&-\frac{ \Gamma(j+5/2)}{ 4^{j+4}\sqrt{\pi}(j+2)!}\left[a_{(A,2)_j}^{(1)}\frac{32(2+j)^2}{(3+2j)(5+2j)}-a_{(A,+)_{j-1}}^{(1)}\frac{64(3+j)^4}{\left(35+24j+4j^2\right)^2}-4a_{(A,+)_{j+1}}^{(1)}\right]\,,
}
where now we must already know $a^{(1)}_{(A,+)_{j\pm1}}$ to determine $a^{(1)}_{(A,2)_{j}}$. Using the formulae for the former in \eqref{AplusFinal} and comparing to \eqref{A22}, we find
\es{A2Final}{
a_{(A,2)_1}^{(1)}&=-\frac{262144}{105} + \frac{212992}{9 \pi^2}\,,\\
a_{(A,2)_3}^{(1)}&=-\frac{16777216}{1617} + \frac{1117782016}{11025 \pi^2}\,,\\
a_{(A,2)_5}^{(1)}&=-\frac{17179869184}{637065} + \frac{47746882994176}{180093375 \pi^2}\,.\\
}

\subsubsection{$\Delta^{(1)}_{(A,0)_{j,0,0}}$ for $j=0,2,\dots,12$ and $\Delta^{(1)}_{(A,0)_{j,1,\overline{q}}}$ for $j=0,2,\dots,10$}
\label{scal}

We will now demonstrate how to compute the sub-leading scaling dimension for the unprotected operator $(A,0)_{j,n,q}$ with twist $2n+2$ and spin $j$. At order $1/c_T$, there are $n+1$ distinct operators of this form, which can be written as double traces of  $\frac12$-BPS operators:
\es{doubleTraceGen}{
[\cO_p \cO_p]_{j,m}=\cO_p\Box^{q}\partial_{\mu_1}\dots\partial_{\mu_j}\cO_p+\dots\,,\qquad \text{for $q=p/2-1\,,p/2+1\dots\,, n$}\,,
}
where $\cO_p$ for $p=2,4,\dots$ are $\frac12$-BPS $(B,+)$ operators in $\mathfrak{so}(8)$ irrep $[00p0]$ with $\Delta=p/2$, as shown in Table \ref{Multiplets}.\footnote{We can also construct double traces of $\cO_p$ for odd $p$, but these do not show up in the stress tensor four-point function.} For instance, $\cO_2\equiv\cO_\text{Stress}$ and  $\cO_4\equiv\cO_{(B,+)}$ in our shorthand notation. In the strict $c_T\to\infty$ limit, all such operators with the same $n$ were indistinguishable and so we could refer to them all by the $p=2$ operator, with scaling dimension \eqref{superMean}. At order $1/c_T$, however, we expect each operators with different $q$ to have different scaling dimensions and OPE coefficients, just like in the maximally supersymmetric AdS$_5$/CFT$_4$ case \cite{Alday:2017xua,Aprile:2017bgs}. For $n>0$, we refer to our results as $\Delta^{(1)}_{(A,0)_{j,n,\overline{q}}}$ to emphasize that they are weighted averages of all $n+1$ operators of this form.

Let us begin with the lowest operator $n=0$ for a given spin $j$, which has $\Delta^{(0)}_{(A,0)_{j,0,0}}=j+2$, i.e twist 2. Since only $(A,0)_{j,n,q}$ operators have anomalous dimensions, when choosing a conformal primary we need only check how many times it appears in $(A,0)_{j,n,q}$. From Table 6 in \cite{Chester:2014fya}, we see that for $\Delta=j+2$ the only unique conformal primary is $(j+4,j+2)_{[0020]}$. We now perform the usual steps of projecting to $Y_{11}$, taking the $s=2$ pole, performing the sum over poles in $t$, extracting $g_{j+4,j+2}^{[0]}(V)$, and then performing the sum over $m$, except we now choose the $U\log U$ coefficient because that is what multiples $\Delta^{(1)}_{(A,0)_{j,0,0}}$ in \eqref{Aexpansion2}. We find
\es{Ascal1}{
\oint_{V=1}\frac{dV}{2\pi i}\frac{F_{-j-2}(1-V)}{(V-1)^{j+3}}\mathcal{A}\big\vert_{U\log U\,Y_{11}}[V]&=\begin{cases}
-\frac{64}{15\pi^2}\qquad\qquad j=0\\
-\frac{16}{105 \pi^2}\qquad\quad \;\;j=2\\
-\frac{32}{3003 \pi^2}\qquad\quad\; j=4\\
-\frac{76}{109395 \pi^2}\quad\quad\;\; j=6\\
-\frac{32}{734825 \pi^2}\quad\quad\;\; j=8\\
-\frac{8}{2982525 \pi^2}\quad\quad \;j=10\\
-\frac{1168}{7125711075 \pi^2}\quad\,\, j=12\\
\end{cases}\,.
}
From the block expansion \eqref{Aexpansion2} and the tables in \cite{Chester:2014fya} we find
\es{Ascal2}{
&\oint_{V=1}\frac{dV}{2\pi i}\frac{F_{-j-2}(1-V)}{(V-1)^{j+3}}\mathcal{A}\big\vert_{U\log U\,Y_{11}}[V]=\\
&\Delta^{(1)}_{j,0,0}\left(\frac{ \Gamma(j+5/2)}{ 4^{j+4}\sqrt{\pi}(j+2)!}\right)\left(\frac{a^{(0)}_{(A,0)_{j,0,0}}}{2}\right)\left(\frac{128 (1 + j)^2 (2 + j)^2}{(1 + 2 j) (3 + 2 j)^2 (5 + 2 j)}\right)\,,
}
where $a^{(0)}_{(A,0)_{j,0,0}}$ are listed for $j=0,2,\dots,12$ in Table \ref{Avalues}. Comparing this to \eqref{Ascal1} we get
\es{AscalFinal}{
\Delta^{(1)}_{(A,0)_{0,0,0}}&=-\frac{1120}{\pi^2}\,,\quad \Delta^{(1)}_{(A,0)_{2,0,0}}=-\frac{2464}{5\pi^2}\,,\quad \Delta^{(1)}_{(A,0)_{4,0,0}}=-\frac{2288}{7\pi^2}\,, \quad \Delta^{(1)}_{(A,0)_{6,0,0}}=-\frac{5168}{21\pi^2}\,,\\
 \Delta^{(1)}_{(A,0)_{8,0,0}}&=-\frac{97888}{495\pi^2}\,,\quad \Delta^{(1)}_{(A,0)_{10,0,0}}=-\frac{165600}{1001\pi^2}\,, \quad \Delta^{(1)}_{(A,0)_{12,0,0}}=-\frac{64728}{455\pi^2}\,,
}
where $\Delta^{(1)}_{0,0,0}$ was already obtained by \cite{Zhou:2017zaw} using the superconformal primary $(2,0)_{[0000]}$.

We now move on to the second lowest twist operators $(A,0)_{j,1,q}$, which has $\Delta^{(0)}_{(A,0,m)_{j,1,q}}=j+4$, i.e. twist 4. While there is no twist 4 conformal primary that only appears in $(A,0)_{j,1,q}$, we choose $(j+6,j+2)_{[0120]}$, because it overlaps with fewer multiplets than other choices. Performing the same first few steps as with $(A,0)_{j,0,0}$, except now choosing the $U^2\,\log U$ coefficient and integrating against $\frac{F_{-j-3}(1-V)}{(V-1)^{j+3}}$, we find
\es{Ascal21}{
\oint_{V=1}\frac{dV}{2\pi i}\frac{F_{-j-3}(1-V)}{(V-1)^{j+3}}\mathcal{A}\big\vert_{U^2\log U\,Y_{11}}[V]&=\begin{cases}
-\frac{128}{75\pi^2}\qquad\qquad\quad\; j=0\\
-\frac{256}{3003 \pi^2}\qquad\quad \quad\;\; j=2\\
-\frac{3904}{853281 \pi^2}\qquad\quad\quad j=4\\
-\frac{20992}{82447365 \pi^2}\quad\quad\quad\; j=6\\
-\frac{15424}{1064761425 \pi^2}\quad\quad\;\; j=8\\
-\frac{63872}{76346904375 \pi^2}\quad\quad \;j=10\\
\end{cases}\,.
}
In the block expansion \eqref{Aexpansion2} we expect to receive contributions from other twist 4 blocks $g^{[0]}_{j+4,j}(V)$, as well as the $k=1$ correction to twist 2 blocks $g^{[1]}_{j'+2,j'}(V)$ for $j'=j,j+2$. Using the explicit formula for these blocks in \eqref{lightconeBlock}, as well as the tables in  \cite{Chester:2014fya}, we get
\es{Ascal22}{
&\oint_{V=1}\frac{dV}{2\pi i}\frac{F_{-j-3}(1-V)}{(V-1)^{j+3}}\mathcal{A}\big\vert_{U^2\log U\,Y_{11}}[V]=\Delta^{(1)}_{(A,0)_{j,1,\overline{q}}}a^{(0)}_{(A,0)_{j,1,q}}\frac{(2+j)(3+j)\Gamma\left(j+\frac12\right)}{4^{j+4}(2j+5)(2j+7)\sqrt{\pi}j!}\\
&+\Delta^{(1)}_{(A,0)_{j,0,0}}a^{(0)}_{(A,0)_{j,0,0}}\frac{2(j+4)!(3j^2+25j+46)\Gamma\left(j+\frac52\right)}{4^{j+2}(6j+3)(2j+3)^2(2j+5)^2(2j+7)(2j+11)\sqrt{\pi}j!^2}\\
&-\Delta^{(1)}_{(A,0)_{j+2,0,0}}a^{(0)}_{(A,0)_{j+2,0,0}}\frac{2(j+3)^2(3j^2+17j+18)\Gamma\left(j+\frac92\right)}{4^{j+3}(6j+9)(2j+5)(2j+7)^2(2j+9)\sqrt{\pi}(j+2)!}\,,
}
where we must already know $\Delta^{(1)}_{(A,0)_{j,0,0}}$ and $\Delta^{(1)}_{(A,0)_{j+2,0,0}}$ to determine $\Delta^{(1)}_{(A,0)_{j,1,\overline{q}}}$. Using the formulae for the former in \eqref{AscalFinal} and comparing to \eqref{Ascal21}, we find
\es{Ascal2Final}{
\Delta^{(1)}_{(A,0)_{0,0,\overline{q}}}&=-\frac{3584}{\pi^2}\,,\qquad \Delta^{(1)}_{(A,0)_{2,0,\overline{q}}}=-\frac{59488}{35\pi^2}\,,\qquad \Delta^{(1)}_{(A,0)_{4,0,\overline{q}}}=-\frac{367744}{315\pi^2}\,, \\
 \Delta^{(1)}_{(A,0)_{6,0,\overline{q}}}&=-\frac{444448}{495\pi^2}\,,\qquad
 \Delta^{(1)}_{(A,0)_{8,0,\overline{q}}}=-\frac{942080}{1287\pi^2}\,,\qquad \Delta^{(1)}_{(A,0)_{10,0,\overline{q}}}=-\frac{619440}{1001\pi^2}\,.
}

\subsection{Comparison to exact results and numerical bootstrap}
\label{comparison}

We now compare these tree level AdS$_4$ supergravity results to CFT$_3$ results. The short operator OPE coefficients $\lambda^2_{(B,2)}$ and $\lambda^2_{(B,+)}$ were computed to all orders in $1/c_T$ in \cite{Agmon:2017xes}. To sub-leading order, the answer is 
\es{exact}{
\lambda^2_{(B,2)}&=\frac{32}{3}-\left(\frac{4096}{9}-\frac{5120}{3\pi^2}\right)c_T^{-1}+O(c_T^{-5/3})\,,\\
\lambda^2_{(B,+)}&=\frac{16}{3}+\left(\frac{1024}{9}+\frac{1024}{3\pi^2}\right)c_T^{-1}+O(c_T^{-5/3})\,,\\
}
which exactly matches the supergravity results \eqref{BplusFinal} and \eqref{B2Final}. 

\begin{figure}[t!]
  \centering
\begin{center}
   \includegraphics[width=0.49\textwidth]{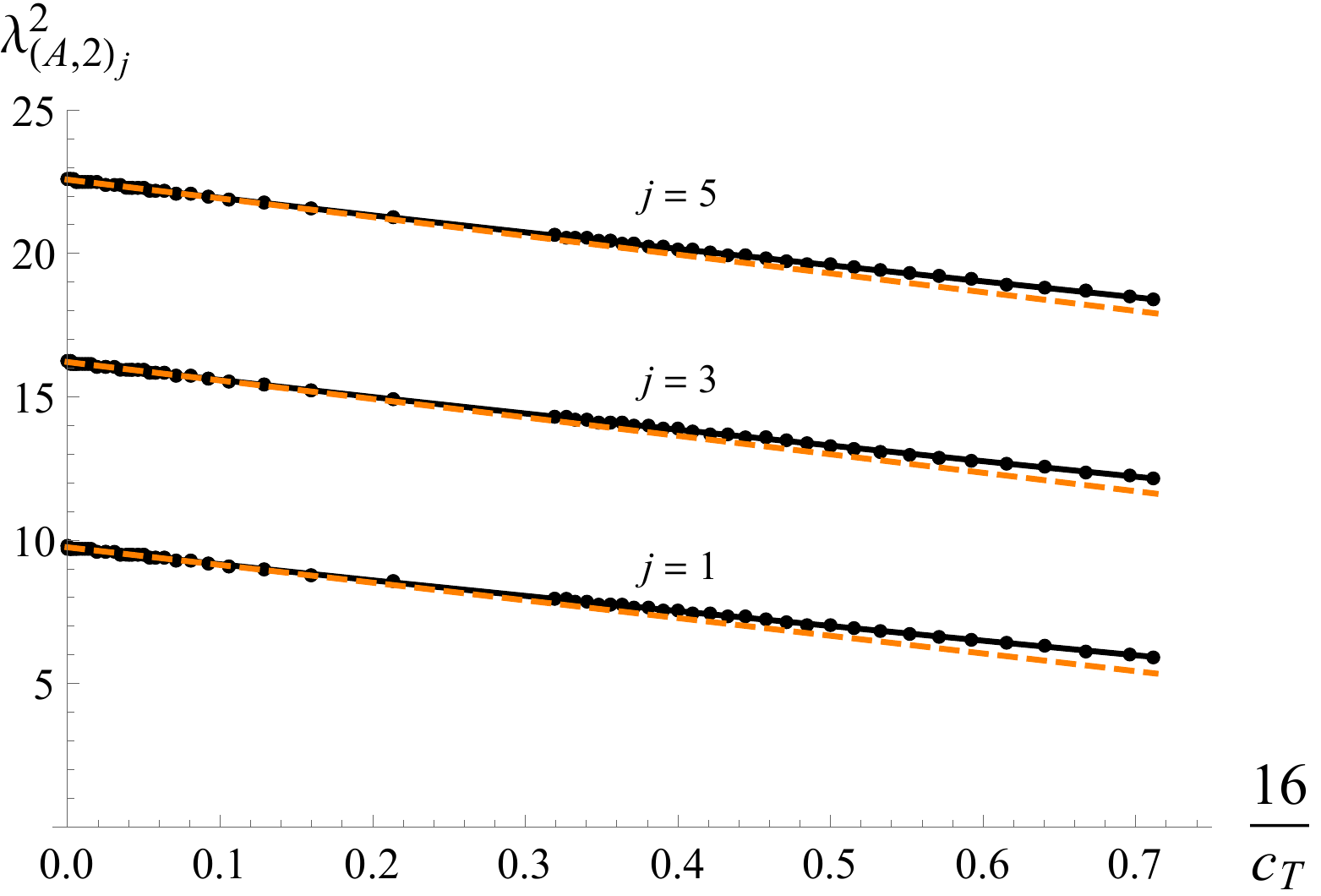}
    \includegraphics[width=0.5\textwidth]{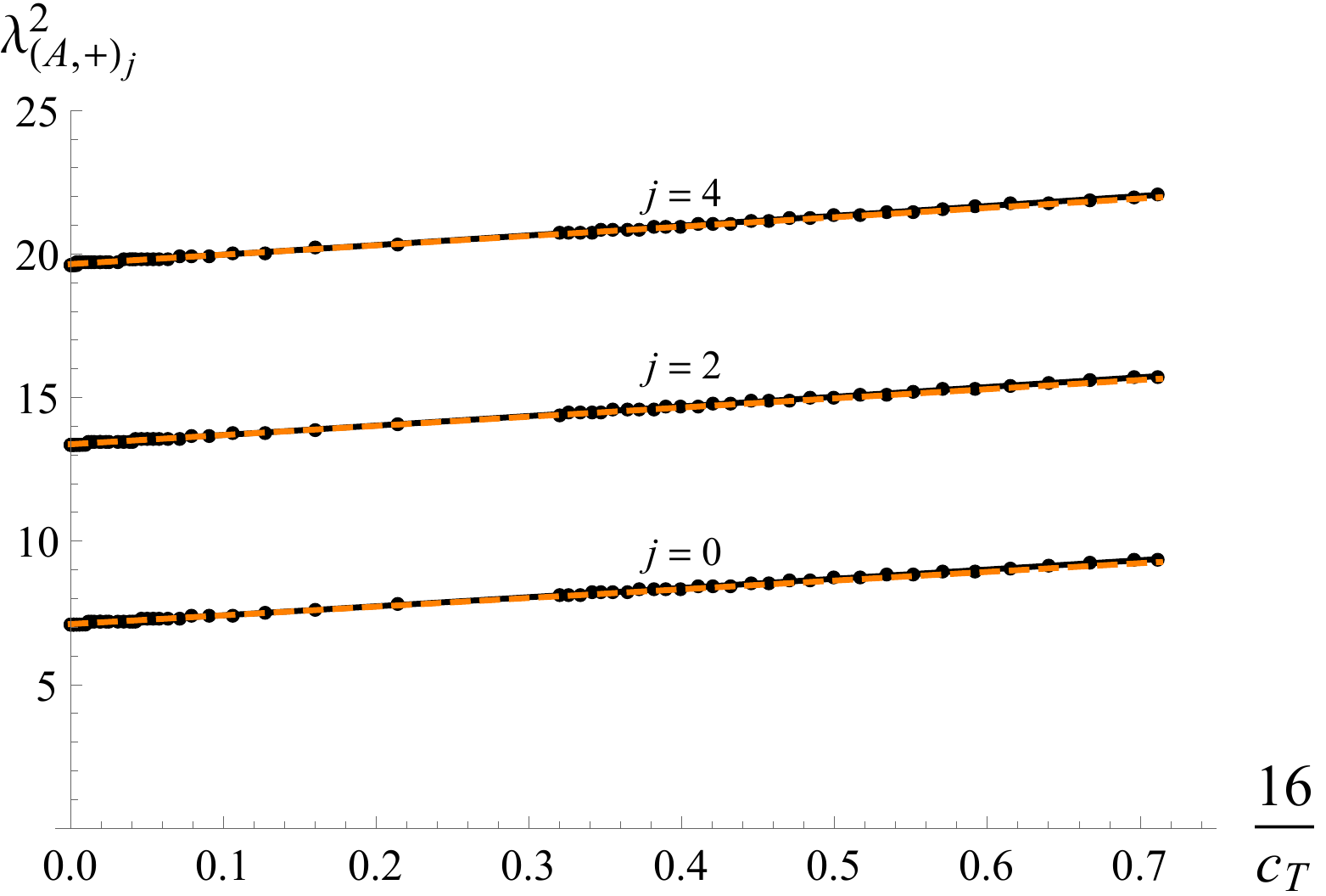}
    \caption{
The $\lambda_{(A,2)_j}^2$ and  $\lambda_{(A,+)_j}^2$ OPE coefficients with spins $j=1,3,5$ and $j=0,2,4$, respectively, in terms of the stress-tensor coefficient $c_T$, where the plot ranges from the generalized free field theory limit $c_T\to\infty$ to the numerical point $\frac{16}{c_T}\approx.71$ where $\lambda_{(B,2)}^2=0$, which is near the lowest interacting theory ABJ$_1$ with $c_T=.75$. The orange dotted lines show the analytic $1/c_T$ corrections  \eqref{A2Final} and \eqref{AplusFinal}.}
\label{APlots}
\end{center}
\end{figure}

There are no exact results for the other operators in $\cO_\text{Stress}\times \cO_\text{Stress}$, but the conformal bootstrap was used to estimate their correction at large $c_T$ in \cite{Agmon:2017xes}. In Table \ref{comparisonTable}, we compare the numerical CFT$_3$ predictions to the analytic AdS$_4$ results computed here. For the semi-short operators $(A,2)_j$ and $(A,+)_j$ and the lowest unprotected operator $(A,0)_{j,0,0}$, we find precise agreement for every value of $j$. In Figure \ref{APlots} we compare the numerical plots of the semi-short OPE coefficients $\lambda^2_{(A,2)_j}$ and $\lambda^2_{(A,+)_j}$ from \cite{Agmon:2017xes} to the exact $1/c_T$ correction \eqref{A2Final} and \eqref{AplusFinal}. The $\lambda^2_{(A,+)_j}$ plots appears to be linear in $1/c_T$, while the $\lambda^2_{(A,2)_j}$ plots depart from linearity for large $1/c_T$. The plots for the other CFT data in \cite{Agmon:2017xes} are not nearly linear, so we do not reproduce them here.

For the second to lowest $(A,0)_{j,1,q}$, we have only been able to compute the average $\Delta^{(1)}_{(A,0)_{j,1,\overline{q}}}$ of the two such operator given in \eqref{doubleTraceGen} for $q=0,1$. The numerical bootstrap was used to compute the anomalous dimension of the lower of these two operators, and so a direct comparison is not possible with this information. Nevertheless, by analogy to the explicit answer for all $n,j,q$ in the AdS$_5$/CFT$_4$ case \cite{Alday:2017xua,Aprile:2017bgs}, we expect that the $q$-dependence is suppressed at large $j$. This expectation is confirmed in Table \ref{comparisonTable}, where we find that $\Delta^{(1)}_{(A,0)_{j,1,\overline{q}}}$ and the bootstrap result are very different for small $j$, but become quite similar for larger $j$, e.g. $j=10$. Note that the bootstrap results for $(A,0)_{j,n,q}$ for $j>4$ are unpublished results computed using the methods of \cite{Agmon:2017xes}, which are being reported here for the first time.

 \begin{table}[htp]
\begin{center}
\begin{tabular}{|c|c|c|}
\hline
 \multicolumn{1}{|c|}{CFT data}  &  ABJ(M) numerical bootstrap & AdS$_4$ Supergravity  \\
  \hline
   $a^{(1)}_{(A,2)_1}$  & $-97$ & $-98.765$\\
  $a^{(1)}_{(A,2)_3}$  & $-102$ & $-102.045$\\
  $a^{(1)}_{(A,2)_5}$  & $-104$ & $-103.470$\\
  $a^{(1)}_{(A,+)_0}$  & $49$& $48.448$\\
  $a^{(1)}_{(A,+)_2}$  & $51$& $51.147$\\
  $a^{(1)}_{(A,+)_4}$ & $52$& $52.155$\\
  $\Delta^{(1)}_{(A,0)_{0,0,0}}$ &  $-109$& $-113.480$\\
  $\Delta^{(1)}_{(A,0)_{2,0,0}}$&  $-49$& $-49.931$\\
  $\Delta^{(1)}_{(A,0)_{4,0,0}}$& $-33$& $-33.118$\\
 $\Delta^{(1)}_{(A,0)_{6,0,0}}$ & $-25$& $-24.935$\\
  $\Delta^{(1)}_{(A,0)_{8,0,0}}$& $-20$& $-20.037$\\
 $\Delta^{(1)}_{(A,0)_{10,0,0}}$ & $-17$& $-16.762$\\
 $\Delta^{(1)}_{(A,0)_{0,1,{q}}}$ &  $-261$& $\overline{-363.135}$\\
  $\Delta^{(1)}_{(A,0)_{2,1,{q}}}$ &  $-145$& $\overline{-172.211}$\\
$\Delta^{(1)}_{(A,0)_{4,1,{q}}}$ &  $-111$& $\overline{-118.287}$\\
$\Delta^{(1)}_{(A,0)_{6,1,{q}}}$ &  $-88$& $\overline{-90.974}$\\
  $\Delta^{(1)}_{(A,0)_{8,1,{q}}}$ &  $-70$& $\overline{-74.167}$\\
  $\Delta^{(1)}_{(A,0)_{10,1,{q}}}$ &  $-60$& $\overline{-62.700}$\\
  \hline
\end{tabular}
\end{center}
\caption{The $1/c_T$ correction to the scaling dimensions $\Delta^{(1)}_{(A,0)_{j,n,q}}$ for the $q=0\,,\dots\,,n$ unprotected operators with spin $j$ and twist $2n+2$, as well as the OPE coefficients squared $a^{(1)}_{(A,+)_j}$ and $a^{(1)}_{(A,2)_j}$ for the semi-short operators of spin $j$, computed from the numerical conformal bootstrap for ABJ(M) in \cite{Agmon:2017xes} and the supergravity calculation in this work. Exact formulae for supergravity are given in \eqref{AplusFinal}, \eqref{A2Final}, \eqref{AscalFinal}, and \eqref{Ascal2Final}. For $n>0$, the exact results refer to averages over $n+1$ distinct operators with the same quantum numbers, as denoted by the overline, while the bootstrap results refers to the lowest of these mixed operators.}\label{comparisonTable}
\end{table}

\section{Conclusion}
\label{conclusion}

In this paper we have developed an efficient algorithm to extract CFT data from Mellin space amplitudes for M-theory on $AdS_4\times S^7$ dual to $\mathcal{N}=8$ SCFT. We then used this algorithm to compute the $1/c_T$ correction to the OPE coefficients of protected operators and the anomalous dimensions of unprotected operators from the tree level Mellin amplitude computed in \cite{Zhou:2017zaw} for the holographic dual of the four-point function of the lowest scalar in the stress-tensor multiplet. This Mellin amplitude was computed using the assumption that the supergravity Lagrangian has a two derivative Einstein-Hilbert kinetic term, and so should be considered an AdS$_4$ gravity calculation. We compared the CFT data extracted from this Mellin amplitude to the same data computed using details of the ABJ(M) theories, and found several remarkable matches. 

For the OPE coefficients of the short $(B,2)$ and $(B,+)$ multiplets, we have exactly matched the tree level supergravity result to the $1/c_T$ term from the all orders in $1/c_T$ formula computed from the protected 1d theory in \cite{Agmon:2017xes}. This formula was derived using the Lagrangian of ABJ(M) theory, and so is an inherently CFT$_3$ result. The match between the AdS$_4$ supergravity and CFT$_3$ results at order $1/c_T$ are a remarkable check of AdS$_4$/CFT$_3$ at the level of local operators at tree level. 

For the other CFT data, we compared the supergravity results to the $\mathcal{N}=8$ numerical bootstrap results of \cite{Agmon:2017xes}, which for large $c_T$ are expected to describe all $\mathcal{N}=8$ ABJ(M) theories. This comparison is summarized in Table \ref{comparisonTable}. The OPE coefficients of the semi-short multiplets $(A,2)_j$ and $(A,+)_j$ and the scaling dimensions for the lowest twist unprotected multiplet $(A,0)_{j,0,0}$ match precisely for all spin $j$. For the second lowest twist $(A,0)_{j,1,q}$ multiplets, we expect there to be two distinct operators $q=0,1$ with these quantum numbers, so we have only been able to compute the average of their scaling dimensions for each $j$. We find that this average converges to the bootstrap prediction for the lowest of these two operators as $j$ increases. This is consistent with the analogous case of maximally supersymmetric AdS$_5$/CFT$_4$, where the dependence on $q$ in the unmixed answer is also subleading in $j$ \cite{Alday:2017xua,Aprile:2017bgs}. The matches we find for this unprotected CFT data, along with the recent calculation of $\Delta^{(1)}_{(A,0)_{0,0,0}}$ in \cite{Zhou:2017zaw}, constitute the first precise check of unprotected quantities in AdS$_4$/CFT$_3$.

Looking ahead, it would also be nice to find a formula for general $j$, $n$, and $q$ for the tree level contributions to double trace operators $(A,0)_{j,n,q}$, as was found for AdS$_5$/CFT$_4$ \cite{Alday:2017xua,Aprile:2017bgs}. In that latter case, the general tree level calculation was a prerequisite for the order $1/c_T^2$ one-loop calculation, and we expect the AdS$_4$/CFT$_3$ case to be similar. To find this general formula, one will likely need to consider more general four-point functions in order to unmix the $(A,0)_{j,n,q}$ operators with $n>0$. Another major barrier to finding such a formula is that unlike the even dimensional blocks, the 3d blocks are not known in closed form, and so the algorithm presented in this work must be implemented order by order in $n$ and $j$ for each $q$, even before we consider mixing. At the very least, it would be worth unmixing the two operators for the $n=1$ case discussed in this work, so that we can precisely compare to the bootstrap results.



\subsection*{Acknowledgments}

I thank S.~Pufu, X.~Zhou, and E. Perlmutter for helpful discussions, and S. Pufu for comments on the manuscript. I am supported in part by the Simons Foundation Grant No 488651 and the Bershadsky Family Scholarship in Science or Engineering.

\appendix
\section{Lightcone blocks}
\label{hog}

In this appendix, we review the results of \cite{Hogervorst:2016hal} that we use to construct the lightcone blocks $g_{\Delta,j}^{[k]}(V)$ in the expansion of the 3d conformal block $G_{\Delta,j}(U,V)$ in \eqref{lightBlocksExp}. 

We begin by defining the 2d global conformal blocks
\es{2d}{
 K_{\Delta,j}(U,V)&=\frac12\left[z^{\frac{\Delta+j}{2}}\bar z^{\frac{\Delta-j}{2}}F_{\frac{\Delta-j}{2}}(\bar z)F_{\frac{\Delta+j}{2}}( z)+\bar z^{\frac{\Delta+j}{2}} z^{\frac{\Delta-j}{2}}F_{\frac{\Delta-j}{2}}( z)F_{\frac{\Delta+j}{2}}(\bar z)\right]\,,\\
 U&\equiv z \bar z\,,\qquad V\equiv(1-z)(z-\bar z)\,,
}
where $F_p(x)$ was defined in \eqref{ortho}. We now decompose $G_{\Delta,j}(U,V)$ into $K_{\Delta,j}(U,V)$ as
\es{3dto2d}{
G_{\Delta,j}(U,V) = \sum_{k=0}^\infty \sum_{j' =j\hspace{-.25cm}\mod2}^jA_{k,j'}(\Delta,j)K^{[k]}_{\Delta+2k,j'}(U,V)\,,
}
where in the conventions defined in \eqref{confBlock} the coefficients $A_{k,j'}(\Delta,j)$ are
\es{Acoeff}{
A_{k,j'}(\Delta,j)=\frac{\pi^{-1}16^{-k} \left(2-\delta_{j',0}\right)\left(\frac12\right)_k\left(\Delta-1\right)_{2k}\left(\frac{\Delta-j'}{2}\right)_k\left(\frac{\Delta+j'}{2}\right)_k\left(\frac{\Delta-j-1}{2}\right)_k\left(\frac{\Delta+j}{2}\right)_kj!}{ k!\left(\frac{j-j'+1}{2}\right)_\frac12\left(\frac{j+j'+1}{2},\right)_\frac12\left(\frac12\right)_j\left(\Delta-\frac{1}{2}\right)_k\left(\frac{\Delta-j'-1}{2}\right)_k\left(\frac{\Delta+j'-1}{2}\right)_k\left(\frac{\Delta-j}{2}\right)_k\left(\frac{\Delta+j+1}{2}\right)_k\left(\Delta+k-1\right)_k }\,.
}
We can now expand \eqref{3dto2d} for small $U$ and compare to \eqref{lightBlocksExp} to read off the lightcone blocks $g_{\Delta,j}^{[k]}(V)$. For instance, for $k=0,1$ the results are given in \eqref{lightconeBlock}.

\bibliographystyle{ssg}
\bibliography{Mellin}

\end{document}